\documentstyle[12pt]{article}
\title{ardo}
\author{Ardo}

\begin{document}                
\baselineskip=23pt
\parindent=20pt
\textfraction{0.1}
\topfraction{0.9}
\bottomfraction{0.9}

\title{Early stage scaling in phase ordering kinetics}
\author{F.Corberi$^{1}$, A.Coniglio$^{1}$ and M.Zannetti$^{2}$}
\maketitle

\noindent
\newline
$1$ Dipartimento di Scienze Fisiche, Universit\`a degli studi di Napoli,
Mostra d'Oltremare, Padiglione 19, 80125 Naples, Italy
\newline
and Istituto Nazionale di Fisica della Materia, Unit\`a di Napoli
\newline
$2$  Dipartimento di Fisica, Universit\`a di Salerno, 84081 Baronissi (SA),
Italy
\newline
and Istituto Nazionale di Fisica della Materia, Unit\`a di Salerno
\date{}
\medskip
\vspace{1 cm}

\begin{abstract}

A global analysis of the scaling behaviour of
a system with a scalar order parameter quenched to zero temperature
is obtained by numerical simulation of the Ginzburg-Landau equation
with conserved and non conserved order parameter.
A rich structure emerges, characterized by early
and asymptotic scaling regimes, separated by a crossover.
The interplay among different dynamical behaviours is
investigated by varying the parameters of the quench and can be interpreted
as due to the competition of different dynamical fixed points.
\end{abstract}
\newcommand {\be} {\begin{equation}}
\newcommand {\ee} {\end{equation}}
\setcounter{page}{1}

\section {Introduction}

In recent years a great effort has been made to understand the
scaling behaviour observed in the
late stage of the phase ordering process following the
temperature quench of a system from an
initial disordered state to a final state inside the
coexistence region, below the critical point. A
theoretical framework for the description of this phenomenon,
well suited both for analytical and
numerical approaches, is provided by the time dependent
Ginzburg-Landau model (TDGL)~\cite{a}.

In the late stage of phase ordering the order parameter
saturates locally to the equilibrium
values, giving rise to configurations with domains
separated by sharp interfaces.
In this regime the only dynamics left in the system is, therefore,
interface motion and the subsequent time
evolution is characterized by the coarsening of domains,
while their morphology remains invariant. This leads to the formulation of
the scaling hypothesis~\cite{b},
whereby the residual time dependence in the system is
due only to the growth of the size of
domains according to the power law $L(t)\sim t^{\frac {1}{z}}$,
with $z=2$ for non conserved
order parameter (NCOP) and $z=3$ in the
conserved case (COP).
According to this hypothesis, the equal time order parameter correlation
function obeys an asymptotic form of
the type
\be
G(|\vec x-\vec x'|,t)\sim L^{-\alpha }(t)\, F
\left [ \frac {|\vec x-\vec x'|}{L(t)}\right ]
\label {1}
\ee
where $\alpha =0$ due to the formation of compact domains
and $F$ is a scaling function. This in turn implies the following
form
\be
C(\vec k,t)\sim L(t)^{d-\alpha}{\cal F}[\vec kL(t)]
\ee
for the structure factor, which is the space Fourier transform of
$G(|\vec x-\vec x'|,t)$.
Although a full theory of scaling is yet to come, the prediction
of the scaling hypothesis for the
correlation function is confirmed by experiments~\cite{c},
numerical simulations~\cite{d} and by exactly
soluble models for NCOP~\cite{e}.
Less is known about the behaviour of quenched systems in
the regime preceeding the asymptotic dynamics~\cite{klein}.
At early times domains are
quickly formed but the order parameter is still away from saturation inside
the ordered regions. As a consequence much more freedom is left to the
system as compared to the asymptotic regime, since modulations of the field are
also possible.

In this paper we present a global analysis of the scaling behaviours
obtained by simulating the TDGL equation
for a two-dimensional system and by varying the
parameters of the Hamiltonian over the $T=0$ manifold of the equilibrium
phase diagram. The time
evolution of the system is followed from the instant of the quench
down to equilibration.
In so doing, we uncover a structure much more rich than usually realized.
In the very early stage, for $t$ smaller than a crossover time $t_1$,
the system relaxes towards equilibrium with a purely
diffusive behaviour.  During this early regime the amplitude
of the order parameter shrinks to zero in order to remove
spatial inhomogeneities and the correlation function obeys
the scaling form~(\ref{1}) with exponents $\alpha =d$ and
$z=2$ or $z=4$ respectively for NCOP or COP. Then, at $t\simeq t_1$,
the system enters an intermediate regime characterized by
exponential growth of the order parameter towards its local
equilibrium value. Eventually, after a characteristic time
$t_2$, the late stage scaling is asymptotically
obeyed, with exponents $\alpha=0$ and $z=2$ or $z=3$ for NCOP or COP.
This whole structure is schematically represented
in fig.1 with symbols to be specified in the following section.
We stress that what we call early stage here precedes the usual early
time behaviour characterized by exponential growth.
For common choices
of the parameters entering the model the crossover time
$t_1$ is too short to make the early stage observable.
However, this
can be greatly amplified by a proper choice of the parameters of the
TDGL equation. For COP this amplification can be obtained by
performing asymmetric (off-critical) quenches.

In the present paper this structure is investigated by
the numerical solution of the TDGL equation and is
interpreted in terms of the interplay between
different fixed points of competing stability.
For short times the dynamics is controlled by
the trivial fixed point of simple diffusion whereas
asymptotically the attractive ordering fixed point dominates.
The intermediate regime corresponds to
the crossover in the time interval
$t_1<t<t_2$ .
This interpretation is supported by the comparison with the
exactly soluble large-$N$ model, where similar
properties are observed analytically~\cite{i}.

The outline of the paper is as follows:
in the next section we introduce the TDGL
equation and set up the notation.
In section 3 we present the results
of numerical simulations of the model in the scalar case
which clearly show the existence of the early scaling regime.
Data are presented for conserved and non-conserved
order parameter, critical and
off-critical quenches. The role of the parameters entering
the TDGL equation and their effect on the early stage
behaviour is elucidated.
Section 4 is dedicated to the comparison with the exact solution
of the large-$N$ model. Finally, in section 5
a summary of the results is presented and concluding remarks
are made.

\section {The model}

We consider a system with a scalar order parameter $\Phi (\vec x , t)$
initially prepared in a configuration $\Phi (\vec x ,0)$
sampled from a high temperature
uncorrelated state with expectations
\be
<\Phi (\vec x ,0)>=0
\label{inicond1}
\ee
and
\be
<\Phi (\vec x,0)\Phi (\vec x',0)>=\Delta \delta (\vec x -\vec x')
\label{inicond2}
\ee
where $\Delta$ is a constant.

For $t\geq 0$ the time
evolution is governed by the Langevin equation
\be
\frac {\partial \Phi (\vec x,t)}{\partial t} = -\Gamma (-i \nabla)^{p}
\frac {\partial {\cal H} [\Phi,\mu]}{\partial \Phi (\vec x,t)}
+\eta  (\vec x,t)
\label {a}
\ee
where $\eta (\vec x,t)$ is the gaussian white noise produced
by the thermal bath at the
temperature of the quench $T$, $\mu$ is the  set of parameters
entering the free energy functional ${\cal H} [\Phi ,\mu]$
and $p=0$ for NCOP while $p=2$ for COP.

In the following we take a free energy functional of the
Ginzburg-Landau form
\be
{\cal H}[ \Phi, \mu]=\int d^d x \left[\frac {R}{2}(\nabla  \Phi)^2 +
\frac {r}{2} \Phi ^2+\frac{g}{4}( \Phi ^2)^2\right]
\label {c}
\ee
with $\mu =(r,g,R)$. The equation of motion then becomes
\be
\frac {\partial \Phi (\vec x,t)}{\partial t}=
-\Gamma (-i \nabla)^{p} \,[-R\nabla ^2 \Phi
+r\Phi +g\Phi ^3] + \eta (\vec{x},t).
\label {d}
\ee
In the study of deep temperature quenches usually one puts $T=0$
(i.e. $\eta (\vec x,t)=0$ in
eq.~(\ref{d})) regarding the temperature as an irrelevant
parameter and $(r,g)$ are chosen in the sector $r<0$, $g>0$
corresponding to
final equilibrium states inside the ordering region.
However, as it will be clear shortly, it is
of considerable interest also the edge of this sector
(i.e. $r=0,\, g\geq 0$), even though no phase
ordering occurs there since the equilibrium value of
the order parameter vanishes.
Thus, the states relevant to our
discussion are those in the $T=0$ plane of the $(T,r,g)$
space with $r\leq 0$ and $g\geq 0$. We will set $R=1$ and $\Gamma=1$
for simplicity.

\section {Numerical results}

As stated in the Introduction, the quantity of interest is
the equal time  order parameter correlation function
\be
G(|\vec x-\vec x'|,t)=<\Phi (\vec x,t)\Phi
(\vec x',t)>-<\Phi (\vec x,t)>^2.
\label {e}
\ee

We consider first the quench to the trivial state with
$\mu _1 \equiv (r=0,\, g=0)$. In this
case eq.~(\ref{d}) can be solved exactly and one finds
\be
C(\vec k,t)=\Delta e^{-2 k^{2+p}t}.
\label{sncop}
\ee
Therefore, defining
$L(t)=(2t)^{1/(2+p)}$ the real space correlation function
is in the scaling form (1) with $\alpha=d, z=2$ for NCOP,
$z=4$ for COP and $F(x)=\exp (-x^{2+p})$.
Notice that in this case eq.(9) is not just an asymptotic
behaviour, but it is obeyed exactly along the whole time
history, from the instant of the quench onward.
In the language of
critical phenomena this means that the width of the critical region
is maximally amplified,
warranting the identification of $\mu _1$ with a (trivial) fixed point
on the $T=0$ manifold.
The next step is the exploration of the domain of attraction of
this fixed point, the search for other
fixed points and for the crossover induced by their competition.

Since for $g\neq 0$ the theory is not soluble we proceed by
numerical simulation.
If scaling holds, from eq.(1) we have $S(t)=G(0,t) \sim L^{-\alpha}(t)
\sim t^{-\alpha/z}$. In the following we will use the behaviours
of $L(t)$ and $S(t)$ to get the pair
of exponents $(\alpha ,z)$.
The evolution of $G(|\vec x-\vec x'|,t)$
is obtained by numerical
solution of eq.~(\ref{d}),
for a two-dimensional 100x100 lattice.
The equal time correlation function is obtained
by averaging over  different realizations of the
time histories (the number of such realizations ranges from 1
to 20, according to the quality of the
data). We have found convenient to extract the
characteristic length $L(t)$ from
the half-height width of $G(|\vec x-\vec x'|,t)$.

We consider first the behaviour of eq.~(\ref{d})
for quenches on the $r=0$ axis.
In this case no phase ordering occurs, in the usual sense,
since eventually the order parameter vanishes.
Nevertheless one can still observe the formation and
subsequent growth of domains which can be defined by
locating their boundary on the contour $\Phi(\vec x,t)=0$.
$L(t)$ can be thought as the typical length
associated to these structures.

The results of our simulations are presented
in fig.2 for NCOP, where the behaviours of $L(t)$ and
of $S(t)$ are displayed
for different values of $g$.
We do not observe significant differences among the behaviours
of $L(t)$ as $g$ is varied and we
conclude that $L(t)$ obeys an asymptotic
growth law with an exponent consistent with $z=2$.
The quantity $S(t)$, on
the other hand, shows a more complex behaviour,
in that the asymptotic scaling
$S(t)\sim t^{-\alpha /z}$ sets in after an initial transient,
which widens as $g$ gets
bigger. For $g=10$ the value of the
exponent $\alpha/z=1.2$ is slightly different from the one found
for $g=0.1$ and $g=1$, where $\alpha/z=1$ implying $\alpha =2$.
This is due the initial transient which for
$g=10$ is not completed over the time of the simulation.

In conclusion the data show an asymptotic scaling behaviour
identical to the one found in the quench to the trivial fixed
point  with $z=2$ and $\alpha =d$. The critical
regime where scaling holds shrinks moving away
from the trivial fixed point along the $r=0$ axis.

The analogous results for COP are presented in fig.$3$.
Again scaling behaviour is
obeyed by $L(t)$ and $S(t)$ with exponents consistent with $z=4$
and $\alpha =2$ but, differently from NCOP,
this feature sets in almost immediately after the quench,
for every value of $g$.
Both for NCOP and COP
quenches to the $r=0$ axis are controlled by the trivial fixed point.
Therefore, all quenches with $r=0$ fall into the same universality
class and the cubic term in eq.~(\ref{d})
is asymptotically irrelevant.

We turn, now, to the phase ordering region by setting
$r<0$ and $g>0$.
When considering quenches inside this region, as it has
been well documented in the literature~\cite{d}, the
asymptotic scaling behaviour is different from
the one just found on the $r=0$ axis. In the phase ordering
region there exists a non trivial
fixed point characterized by
($z=2,\, \alpha =0$) for NCOP and ($z=3,\, \alpha =0$) for COP, whose
domain of attraction is the entire sector ($r<0,\, g>0$).
For quenches sufficiently close to
$r=0$ axis, but inside the ordering region, we expect to
observe the crossover from trivial
to non trivial scaling behaviours.

We have performed simulations with $g=1$ and three
different values of $r$, both for NCOP
(fig.$4$) and COP (fig.$5$). For NCOP, $L(t)$ scales with the
same exponent $z=2$ both when the
dynamics is dominated by the fixed point of the ordered region
or by the trivial one.
More interesting is the behaviour of $S(t)$, which is characterized
by the sequence of three regimes.
In the early regime, whose duration is shorter the farther away from
the $r=0$ axis the quench
occurs, we find that $S(t)$ decreases with
a behaviour similar to the one found in the case of a quench
on the $r=0$ axis.
During this early regime the order parameter evolves as if
the potential did not have a
double well and relaxes locally toward zero in order to
eliminate spatial inhomogeneities as in the
quenches to the trivial fixed point. This occurs because, as a consequence of
the initial disordered state, the gradient term dominates in eq.~(\ref{c}).
Then, at some time $t_1$ (see the schematic representation
of fig.1), the double well structure starts to play a role.
At this point $S(t)$ stops decreasing and enters
the intermediate time regime characterized by
exponential growth of the order parameter toward local equilibrium
at the bottom of the wells. In this regime $L(t)$ is approximately
constant, as in the linear theories~\cite{h}.
The intermediate regime terminates
with the formation of domains within which the order parameter
is close to saturation.
{}From this point onward the late stage is entered with
dynamics dominated by interface motion.
Since the system now is close to saturation
$S(t)\sim S(\infty)= -\frac {r}{g}$, implying scaling
controlled by the non trivial fixed point with $\alpha =0$.

The case of COP, shown in fig.5, is particularly interesting
because this whole crossover
structure is manifested also in the growth law of
the size of domains $L(t)$.
In the first regime, whose duration
is again controlled by the distance from the $r=0$ axis,
the dynamics is dominated by the
trivial fixed point, with scaling exponents $z=4$ and $\alpha =d$.
As stressed before for quenches on the $r=0$ axis, no
initial transient is observed for COP.
This allows a precise
determination of the exponents $\alpha$ and $z$ in the early regime
obtaining $\alpha=2$ and $z=4$.
Then, when the system feels the presence of the local potential,
the exponential growth of $S(t)$
is accompanied by an approximately constant behaviour of $L(t)$
(or, at least, by a lower growth rate) which
characterizes the intermediate time regime. Later on, when $S(t)$
saturates to its asymptotic value, the system enters the
late stage, dominated by the non-trivial
fixed point with $\alpha =0$ and $z=3$
(this is not shown in fig.$5$
since the $z$ exponent reaches its asymptotic value for much longer times).

Finally we consider the effect of performing off-critical quenches, in the
case of a conserved order parameter, by
varying $m=<\Phi (\vec x,t)>$ through the region $m^2 < -\frac {r}{g}$
(see fig.6).
The asymptotic scaling
behaviour inside this region is still controlled~\cite{cop}
by the same non trivial fixed point of the symmetric quenches,
with $z=3$ and $\alpha =0$. The
domain of attraction of this fixed point is, therefore,
the entire region $m^2 <-\frac{r}{g}$
where phase separation occurs. At the
intersection of the coexistence curve with the
$T=0$ manifold (i.e. at the very edge of this region, with
$m^2=-\frac {r}{g}$) no phase
separation occurs since the field reaches a uniform configuration.
In this case the amplitude of $G(|\vec x-\vec x'|,t)$, namely $S(t)$,
vanishes asymptotically approaching equilibrium, as for a critical quench on
the $r=0$ axis.
More precisely in this point of the phase diagram
the long time behaviour of the model is no
longer controlled by the same fixed point as for critical quenches,
which on the contrary requires
$S(t)\to$ constant, and we expect to observe a dynamics
reminiscent of the $r=0$, $m=0$ case, where also $S(t)
\rightarrow 0$.
Therefore it is of considerable interest to understand the nature of the
quenches close to $m^2=-\frac {r}{g}$ and
to study the crossover phenomena induced on the dynamics
of the system for $m^2 <-\frac {r}{g}$.

In order to address these questions we have simulated asymmetric
quenches by preparing the initial
configurations with different values of $m^2$ in the range
(0,$-\frac {r}{g}$).
First of all it must be stressed that, for quenches inside the metastability
region, i.e. for $m^2$
sufficiently close but different from $-\frac {r}{g}$ (see fig.6),
the influence of the ordering
fixed point with $(\alpha =0$ and $z=3)$ is hardly observed,
because the system relaxes into a metastable state inside a single minimum of
the local potential (the one of the majority phase). This can be avoided by
increasing the variance $\Delta$ of the initial condition.
When stable equilibrium is reached in the final state,
the behaviour of $S(t)$ reveals again the
presence of three distinct dynamical regimes. In fig.7
this quantity is shown for quenches to $(r=-1,\, g=1)$ with
different values of $m$. In the
early regime, whose duration increases indefinitely
approaching the coexistence curve, we again
find the trivial scaling behaviour with $(\alpha =d,z=4)$
as for $r=0$ and $m=0$.
Later on the system enters the
intermediate and eventually  the asymptotic regime,
with the same features described in the case of
critical quenches. In the renormalization group language this
means that, by increasing $m$ from 0
toward the coexistence line,
the trajectories start closer to the domain of attraction of
the trivial fixed point $r=0$, $m=0$. This is analogous to
the amplification of the early regime in the critical
quench by decreasing the value of $r$.
However, asymptotically, the non-trivial fixed
point prevails for $m^2<-\frac {r}{g}$.
Therefore  in the space of the parameters
($r,\, g,\, m$) the region $m^2=-\frac {r}{g}$ and the axis $r=0$
axis are the domain
of attraction of the trivial fixed point.

We conclude this section observing that, as far as the role
of the parameter $R$ in eq.~(\ref{d}) is concerned, by rescaling one can
easily show that the effect of increasing $R$ corresponds
to a magnification of the space and time scales.
Since $R$ controls the range of the interaction
it is conceivable that the early stage scaling regime
is more clearly observable in systems with sufficiently long
range interactions.

\section {Comparison with the large-$N$ model}

In this section we compare the results of our
simulations with the solution of the large-$N$ model~\cite{i}.
When the order parameter is an $N$-component
vector field $\vec{\Phi}(\vec{x})=(\Phi_{1}(\vec{x}),
...,\Phi_{N}(\vec{x}))$ in the limit of an infinite
number of components eq.(7) is linearized
\be
\frac {\partial \vec \Phi (\vec x,t)}{\partial t}=-(i\nabla)^{p}
\left [ -\nabla ^2 +r+gS(t)\right ] \vec \Phi (\vec x,t)
\label{ninf}
\ee
where $S(t)$ must be computed self-consistently
through $S(t)=<\vec{\Phi}^{2}(\vec{x},t)>$.
{}From the exact solution of this model
one finds that the asymptotic scaling properties
depend as in the scalar case on the
pair of coupling constants $r$ and $g$.
More specifically one finds that there is a universality class,
under each heading NCOP or COP, for each of the following
three regions in the ($r,g$) space

$\mu_1 \equiv (r=0,g=0)$ (trivial critical state)

$\mu_2 \equiv (r=0,g>0)$ (non trivial critical states)

$\mu_3 \equiv (r<0,g>0)$ (phase ordering region).

In the renormalization group language this means that there are three
fixed points and the extension of the universality classes
depends on the relative stability of these fixed points.
Actually, for quenches to $\mu_2$ there is
a critical dimensionality $d_c$, which
depends on the initial condition, above which the non
linearity in the problem becomes irrelevant.
For an initial condition of the type
\be
C(\vec k,t=0)=\frac{\Delta}{k^\theta}
\ee
one finds $d_c=d_{l}+\theta$, where $d_{l}$
is the lower critical dimensionality of the static problem.
With $\theta=0$
one has $d_{c}=d_{l}=2$.
Therefore, solving the model for $d>2$ one finds the following
asymptotic behaviours for NCOP

\be
\mu _1 ) \mbox{    } C(\vec k,t) \sim F(x)
\ee
\be
\mu _2 ) \mbox{    } C(\vec k,t) \sim F(x)
\label{mu2n}
\ee
\be
\mu _3 ) \mbox{    } C(\vec k,t) \sim  L^d(t)F(x)
\label{mu3n}
\ee
with $L(t)\sim t^{1/2}$ and $F(x)=e^{-x^2}$.

Let us briefly comment on these results. First of all
we have $z=2$ everywhere.
For quenches to the trivial state $\mu_1$ the large-$N$ model
and the scalar model coincide since the local potential is absent.
The same expression (9) and (12) is found for the structure factor.
For quenches to $\mu_2$ the long time behaviour is the same as for the
quench to $\mu_1$, up to corrections to scaling,
as expected since $d>d_c$. The scaling behaviour is
characterized by $(\alpha=d,z=2)$ both for quenches
to $\mu_{1}$ and $\mu_{2}$.

Different is the case of quenches
inside the region of coexisting phases, where eq.~(\ref{mu3n})
is asymptotically obeyed in any dimension, showing that
no upper critical dimensionality exists and therefore
that the non linearity of the problem is always relevant.
In this case scaling is characterized by $(\alpha=0,
z=2)$. Furthermore, computing the early time behaviour
for $|r|$ sufficiently small a scaling
regime is found which is identical to the one found
for quenches to $\mu_1$. This behaviour is separated
from the asymptotic scaling regime described above
by an intermediate regime of exponential growth.

In the large-$N$ model the variety of asymptotic
properties is more complex
when quenches with COP are considered, due to the
existence of multiscaling. The following behaviours are found
\be
\mu_1 ) \mbox{    } C(\vec k,t) \sim F(x)
\ee
\be
\mu_2 ) \mbox{    } C(\vec k,t) \sim F(x)
\ee
\be
\mu_3 ) \mbox{    } C(\vec k,t) \sim L(t)^{\alpha(x)}
\label{mu3c}
\ee
where $L(t)\sim t^{1/4}$, $F(x)=e^{-(x^4+cx^2)}$ and
$\alpha (x)=d[1-(x^2-1)^2]$ is a generalized $x$-dependent
exponent. The general discussion of fixed points and their
relative stability goes along the same lines as for NCOP.

Now we make a qualitative comparison with
the results for the scalar case presented in the previous section.
The first observation concerns the existence of an
early stage scaling regime
which is observed in both models for quenches in the
phase ordering region, with the same exponents $\alpha$
and $z$. This was to be expected since the very nature of this
behaviour is due to the trivial fixed point
with $r=g=0$, where the two models coincide. We emphasize that
the occurrence of early scaling is particularly interesting for
conserved scalar
order parameter since in this case, differently from the
large-$N$ limit, a crossover from $z=4$ to $z=3$
exists in the growth law. Secondly, we comment on the existence
of a critical dimensionality for quenches to $\mu_{2}$. In
the large-$N$ model $d_{c}=d_{l}=2$. In the scalar case
$d_{l}=1$. However, we don't know whether $d_{c}$ is still
equal to $d_{l}$. The results of section 3 indicate that
$d=2$ is above the critical dimensionality.
For what concerns the late stage regime, the two models are quite
similar for NCOP, since scaling is obeyed with
the same exponents. With COP, on the contrary, there
is less of an analogy since the vector model exhibits multiscaling.

\section {Conclusions}

In this paper we have analyzed the quench to zero temperature
of a system with a scalar order parameter through the numerical solution
of the time dependent Ginzburg-Landau equation, both with
conserved and non conserved order parameter. We have paid particular
attention to the early stage of the quench. According to the
values of the parameters $(r,g)$ which characterize the final equilibrium
state, an early stage scaling regime associated to pure diffusive
behaviour may be observed before the usual late stage scaling regime sets in.
This is a crossover pattern of the same type found in the analytical
solution of the large-$N$ model. In that case the pure diffusive behaviour
is associated to the trivial fixed point $(r=0,g=0)$
while the late stage scaling behaviour is associated with the non trivial fixed
point $(r<0,g>0)$ of phase ordering. The competition between
the two fixed points determines the crossover pattern. The question of
the observability of this phenomenon is clearly related
to the time span of the early stage. In the paper we have
mentioned long range forces and off critical quenches toward the
coexistence curve as means of amplification of the early stage.

\newpage

\section{Figure captions}

Fig.1 - Schematic representation of the dynamical response of the scalar
model, both for NCOP and COP. The three regimes are shown, separated by the
crossover times $t_1$ and $t_2$.

Fig.2 - Behaviour of $L(t)$ (fig.2a) and of $S(t)$ (fig.2b)
for a non conserved system
quenched on the $r=0$ axis, with different values of $g$ ($g=0.1$,
$g=1$ and $g=10$). The continuous lines represent
respectively the power laws $t^{\frac {1}{2}}$ and $t^{-1}$ associated to
trivial scaling. Best fit yields $z=2.01\pm 0.05$ for every value of $g$
and $\alpha /z=1.0\pm 0.1$ for $g=0.1$ and $g=1$, $\alpha /z=1.2 \pm 0.1$
for $g=10$.

Fig.3 - Behaviour of $L(t)$ (fig.3a) and of $S(t)$ (fig.3b)
for a conserved system
quenched on the $r=0$ axis, with different values of $g$ ($g=0.1$,
$g=1$ and $g=10$). The continuous lines represent
respectively the power laws $t^{\frac {1}{4}}$ and $t^{-\frac{1}{2}}$
associated to
trivial scaling. Best fit yields $z=3.98\pm 0.05$
and $\alpha/z=0.51\pm 0.03$,
for every value of $g$.

Fig.4 - Behaviour of $S(t)$ for a non conserved system
quenched inside the phase consistence region, with $g=1$ and different
values of $r$ ($r=-0.005$,
$r=-0.05$ and $r=-1$). The duration of the early scaling
regime increases as $|r|$ decreases.

Fig.5 - Behaviour of $L(t)$ (fig.5a) and of $S(t)$ (fig.5b)
for a conserved system
quenched inside the phase consistence region, with $g=1$ and different
values of $r$ ($r=-0.03$, $r=-0.3$ and $r=-1.3$).
The continuous lines represent the trivial scaling behaviour with $z=4$
and $\alpha =2$. Best fit of data with
$r=-0.03$ yields $z=4.1\pm 0.1$ and $\alpha/z=0.49\pm 0.05$.

Fig.6 - The phase diagram in the (m,T) plane.
The phase coexistence region and the spinodal line, above which metastability
occurs, are shown.

Fig.7 - Behaviour of $S(t)$ for a
conserved system with $r=-1$, $g=1$ and $m$ ranging between
$0$ and $0.7$ (from the top to the bottom: $m=0$, $m=0.1$, $m=0.2$, $m=0.3$,
$m=0.4$, $m=0.5$, $m=0.6$ and $m=0.7$). The lower straight line
represents trivial scaling with $\alpha /z=0.5$.

\end{document}